\documentclass[a4paper]{PoS}
\usepackage{graphicx}

\def\cp{$CP$\/}

\def\ra{\!\rightarrow\!}
\def\dbar{\overline{D}{}^{\,0}}

\def\simge{\mathrel{%
   \rlap{\raise 0.511ex \hbox{$>$}}{\lower 0.511ex \hbox{$\sim$}}}}
\def\simle{\mathrel{
   \rlap{\raise 0.511ex \hbox{$<$}}{\lower 0.511ex \hbox{$\sim$}}}}

\title{
\vskip-1.0in
\begin{flushright}
\normalsize{\rm
UCHEP--17--01 \\
19 January 2017
}
\end{flushright}
\vskip1.0in
Charm Physics Prospects at Belle II
}

\ShortTitle{Charm Physics Prospects at Belle II}

\author{\speaker{A. J. Schwartz} \\
        Physics Department, University of Cincinnati, Cincinnati, Ohio 45221 USA \\
        E-mail: \email{alan.j.schwartz@uc.edu}
       \vskip0.15in
        {\rm (on behalf of the Belle II Collaboration)}
}

\abstract{The Belle II experiment is under construction at the
KEK laboratory in Japan. Belle II will study $e^+e^-$ collisions
at or near the $\Upsilon(4S)$ resonance with the goal of collecting
50~ab$^{-1}$ of data, which is a large increase over that recorded by 
the Belle and BaBar experiments. This data will provide a large 
sample of charm meson decays, and Belle II will have a very active 
charm physics program. Here we discuss some highlights of this 
program, focusing on measurements of mixing, \cp\ violation, 
and leptonic decays.}

\FullConference{VIII International Workshop On Charm Physics\\
		5-9 September, 2016\\
		Bologna, Italy}

\begin{document}

\section{Introduction}

There are many strategies to search for new physics beyond 
the Standard Model. Besides studying $B$ meson decays,
for which the Belle, BaBar, and LHCb experiments were designed, one
can study $D$ meson decays, and all three experiments have produced
numerous results in this area. While a hadron experiment such as LHCb
produces a much larger sample of charmed mesons and baryons than does
an $e^+e^-$ experiment such as Belle, studying charm decays at an 
$e^+e^-$ experiment has advantages. For example, an $e^+e^-$ 
experiment has lower backgrounds, higher trigger efficiency, excellent 
$\gamma$ and $\pi^0$ reconstruction (and thus $\eta$, $\eta'$, and 
$\rho^+$ reconstruction), high flavor-tagging efficiency with low 
dilution, and numerous control samples with which to study systematics.
Due to good detector hermeticity and knowledge of the initial state
energy and momentum, missing energy and ``missing mass'' analyses are 
straightforward. In addition, absolute branching fractions can 
be measured. 

The Belle experiment at the KEK laboratory in Japan is now
being upgraded to the Belle~II experiment. The goal of Belle II 
is to collect a data set corresponding to 50 times what Belle obtained. 
Here we discuss the charm physics prospects of this large data 
set, focusing on measurements of charm mixing, \cp\ violation, 
and leptonic decays.

\section{\boldmath Mixing and \cp\ Violation}

One emphasis of the Belle II charm physics program will
be to make high precision measurements of charm mixing and 
search for \cp\ violation. These measurements typically depend 
on measuring the decay time of $D^0$ mesons, and, due to 
an improved vertex detector,  the decay time resolution of
Belle~II is expected to be superior to that of Belle and 
BaBar. Whereas Belle used a four-layer silicon-strip detector, 
Belle II will use four layers of silicon strips 
plus two layers of silicon pixels, with the
smallest pixel size being $55\times 50$~$\mu$m$^2$. 
These pixel layers will lie only 14~mm and 22~mm from 
the interaction point (IP).

\begin{figure}[htb]
\hbox{
\includegraphics[width=0.34\textwidth,angle=-90]{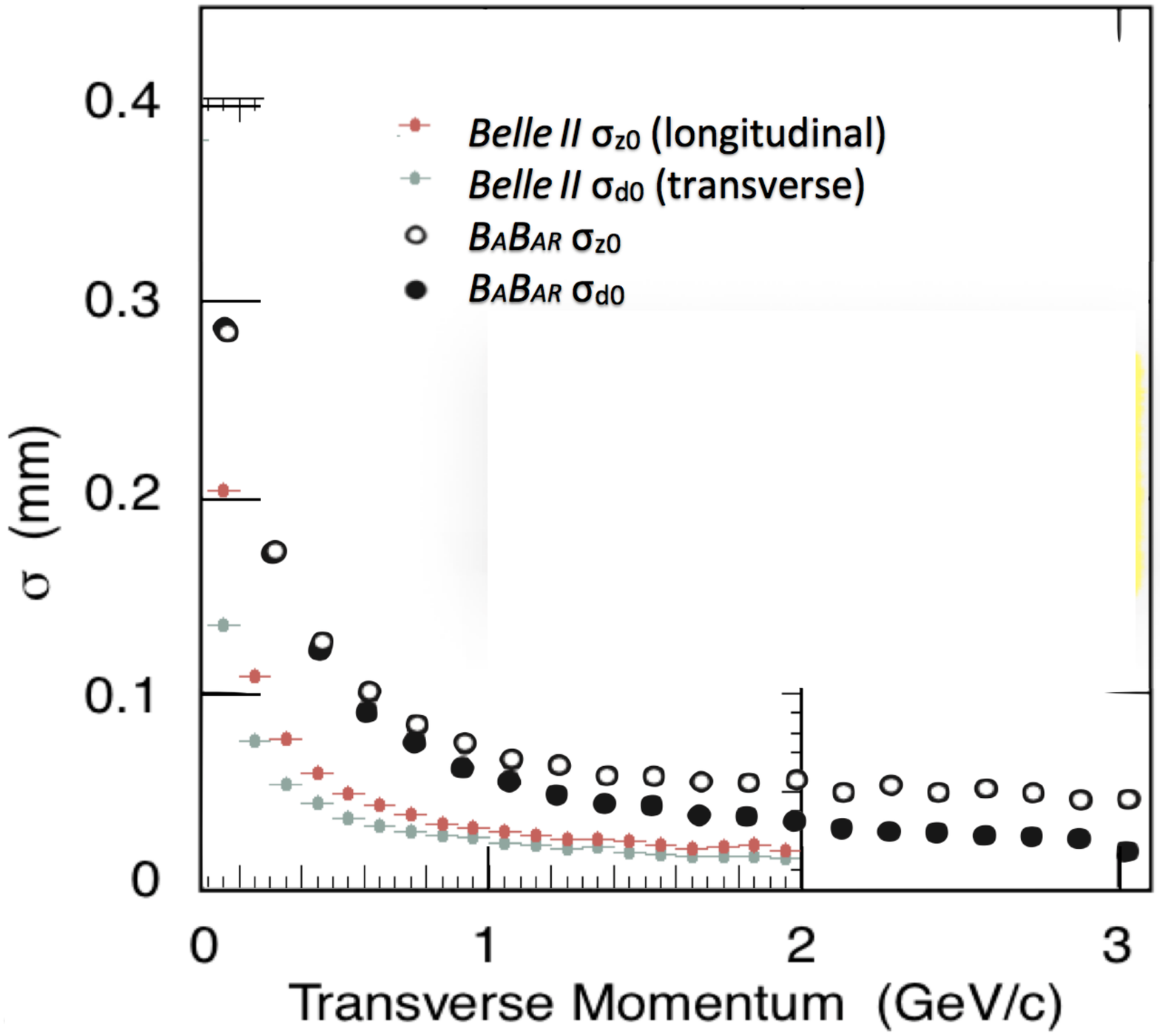}
\hskip0.40in
\includegraphics[width=0.31\textwidth,angle=-90]{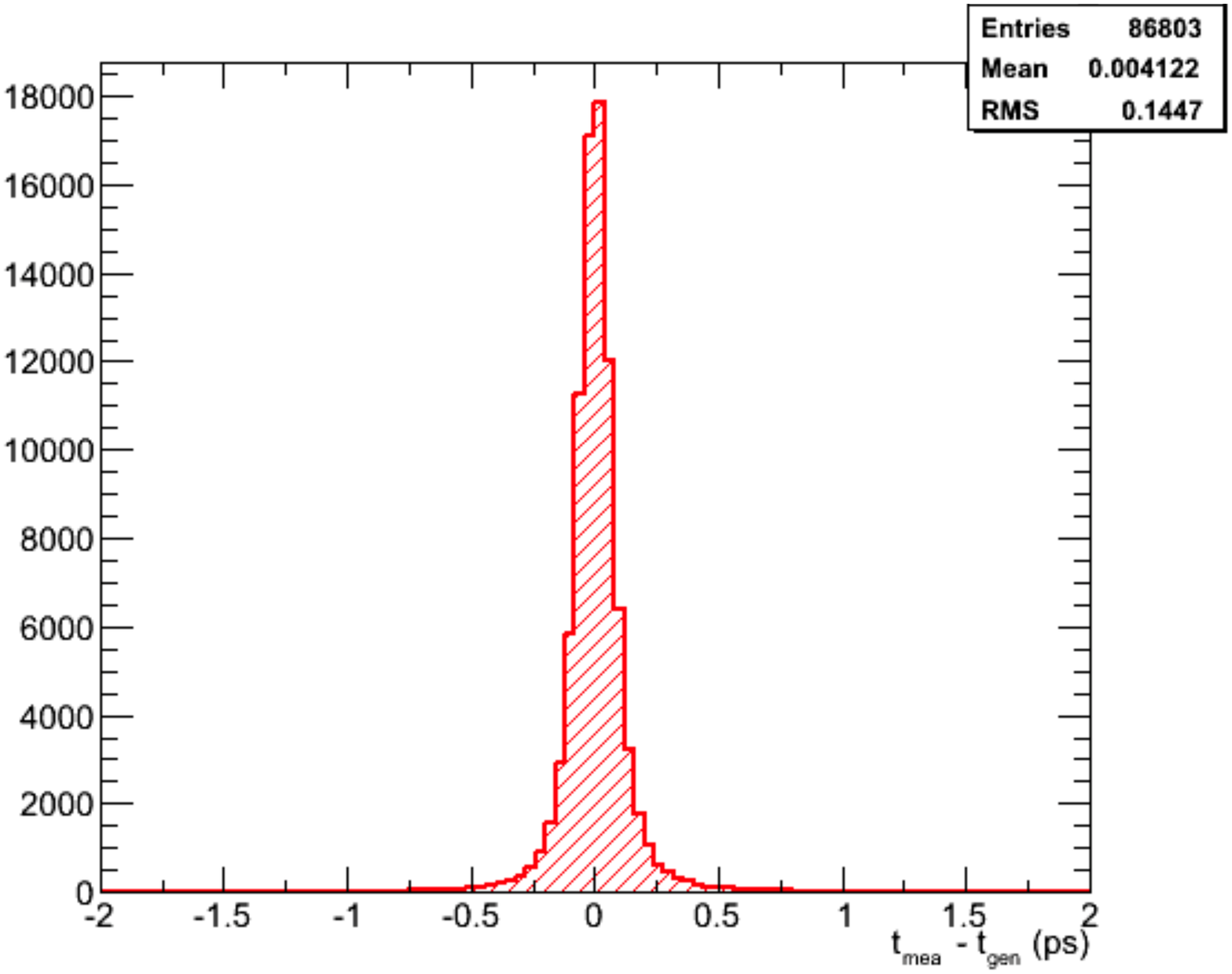}
}
\caption{
Left: Belle II track impact parameter with respect to the IP as 
a function of track momentum, as obtained from MC simulation.
Right: Belle II residuals of decay vertex position, from MC
simulation. The RMS of this distribution is 140~fs.}
\label{fig:resolution}
\end{figure}

Figure~\ref{fig:resolution}~(left)
shows the resolution on track impact parameter with respect to the
IP as a function of track momentum, as obtained from Monte Carlo (MC) 
simulation. Also shown are corresponding results from BaBar.
The figure illustrates 
that the Belle II resolution should be half that of BaBar. 
Figure~\ref{fig:resolution}~(right) plots the residuals of decay 
time for a large sample of MC $D^0\ra\pi^+\pi^-$ decays. The RMS 
of this distribution is 140~fs, which is about half the decay 
time resolution of Babar (270~fs).

To study the sensitivity of Belle II to $D^0$-$\dbar$ mixing 
parameters $x=\Delta M/\Gamma$ and $y=\Delta\Gamma/2\Gamma$,
and to \cp-violating parameters $|q/p|$ and ${\rm Arg}(q/p)=\phi$, 
two studies have been performed. The first study uses ``toy MC'' 
to generate $D^0\ra K^+\pi^-$ decays~\cite{charge-conjugates},
and their decay time distribution is fitted. The second study 
uses a time-dependent Dalitz generator based on 
EVTGEN~\cite{EVTGEN} to generate multi-body $D^0\ra K^+\pi^-\pi^0$ 
decays, and their time-dependent Dalitz distribution is fitted.

\subsection{\boldmath $D^0\ra K^+\pi^-$ Decays}

For this study we generate an ensemble of 1000 ``toy'' MC 
experiments, with each experiment consisting of a sample of 
$D^0$ decays and a separate sample of $\dbar$ decays. The number 
of decays in each sample corresponds to 5~ab$^{-1}$, 20~ab$^{-1}$, 
and 50~ab$^{-1}$ of data. For the full dataset (50~ab$^{-1}$), there
are 438600 $D^0\ra K^+\pi^-$ decays. After generation, the decay 
times are smeared by the expected decay time resolution 
of 140~fs, and the resulting $D^0$ and $\dbar$ time 
distributions are simultaneously fitted. Backgrounds are not
yet included in this study; however, a first look at backgrounds 
indicates that when backgrounds are included, the fitted errors 
increase by only a small amount. 

The probability density functions used for the fit are the 
convolution of the following theoretical expressions with 
Gaussian resolution functions:
\begin{eqnarray}
D^0(t) & = & e^{-\Gamma t}\left\{
R^{}_D + \left|\frac{q}{p}\right|\sqrt{R^{}_D}(y'\cos\phi - x'\sin\phi)(\Gamma t) +
\left|\frac{q}{p}\right|^2\frac{x'^2+y'^2}{4} (\Gamma t)^2\right\} \\
\dbar(t) & = & e^{-\Gamma t}\left\{
\overline{R}^{}_D + \left|\frac{p}{q}\right|
\sqrt{\overline{R}^{}_D}(y'\cos\phi + x'\sin\phi)(\Gamma t) +
\left|\frac{p}{q}\right|^2\frac{x'^2+y'^2}{4} (\Gamma t)^2\right\}\,,
\label{eqn:dkp_decay_time}
\end{eqnarray}
where $x' = x\cos\delta + y\sin\delta$,  
$y' = -x\sin\delta + y\cos\delta$, and 
$\delta$ is the strong phase difference between 
$\dbar\ra K^-\pi^+$ and $D^0\ra K^-\pi^+$ amplitudes.
The parameter $R^{}_D$ ($\overline{R}^{}_D$) is the ratio
of amplitudes squared
$|{\cal A}(D^0\ra K^+\pi^-)/{\cal A}(D^0\ra K^-\pi^+)|^2$
($|{\cal A}(\dbar\ra K^-\pi^+)/{\cal A}(\dbar\ra K^+\pi^-)|^2$).
The preliminary results are listed in Table~\ref{tab:dkp_results}.
The final Belle II sensitivity for $|q/p|$ is~$<0.1\%$, and that 
for $\phi$ is $6^\circ$. These results are a significant 
improvement over that which Belle and BaBar achieved.

\begin{table}[hbt]
\begin{center}
\begin{tabular}{l|lll}  
Parameter &  5~ab$^{-1}$ &  20~ab$^{-1}$ &  50~ab$^{-1}$ \\
\hline
$\delta x'$ (\%) & 0.37 & 0.23 & 0.15 \\
$\delta y'$ (\%) & 0.26 & 0.17 & 0.10 \\
$\delta |q/p|$ (\%) & 0.20 & 0.09 & 0.05 \\
$\delta \phi$ ($^\circ$) & 16 & 9.2 & 5.7 \\
\hline
\end{tabular}
\caption{Preliminary results of a toy MC study of $D^0\ra K^+\pi^-$ 
decays: uncertainty on mixing parameters $x'$ and $y'$, and on
\cp-violating parameters $|q/p|$ and $\phi$, for three values 
of integrated luminosity.}
\label{tab:dkp_results}
\end{center}
\end{table}

\subsection{\boldmath $D^0\ra K^+\pi^-\pi^0$ Decays}

For this study we generate an ensemble of 10 experiments using
a time-dependent Dalitz generator based on EVTGEN~\cite{EVTGEN}. 
Each experiment consists of 225000 $D^0\ra K^+\pi^-\pi^0$ decays, 
corresponding to 50~ab$^{-1}$ of data. The decay times are smeared 
by a resolution of 140~fs, and the decay model used to generate
and fit the Dalitz 
plot is the isobar model used by BaBar~\cite{babar_dkpp_dalitz}.
Possible \cp\ violation and backgrounds are neglected in this phase
of the study. More details are given in Ref.~\cite{longki_CPC}.

\begin{figure}[htb]
\vskip-1.40in
\includegraphics[width=0.78\textwidth,angle=-90]{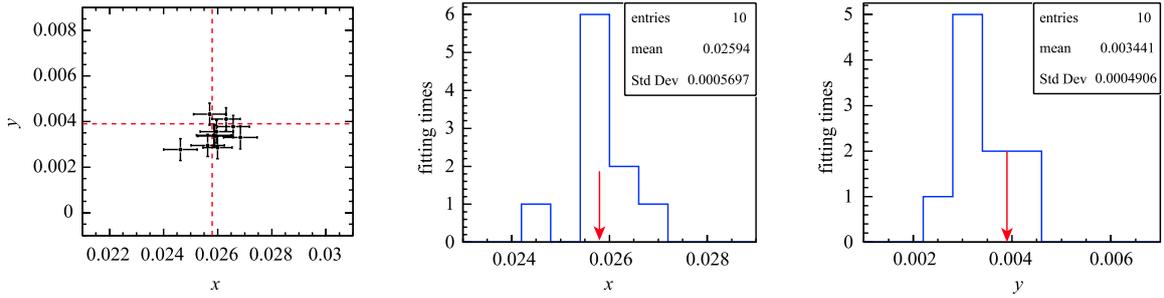}
\vskip-1.60in
\caption{
Left: preliminary results of fitting an ensemble of ten 
experiments, with each experiment corresponding to 50~ab$^{-1}$ 
of data~\cite{longki_CPC}.
Middle and right: projections of the left-most plot. The RMS 
values of these distributions are $\delta x = 0.057\%$ and 
$\delta y = 0.049\%$.}
\label{fig:kpp_dalitz_results}
\end{figure}

\begin{figure}[htb]
\vskip-1.60in
\includegraphics[width=0.78\textwidth,angle=-90]{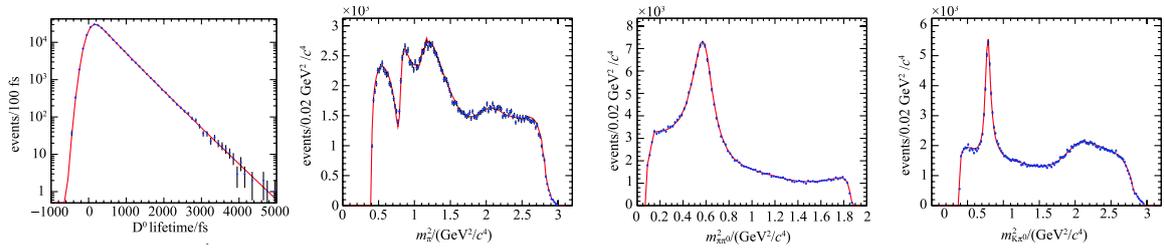}
\vskip-1.74in
\caption{
Projections of the fit to $D^0\ra K^+\pi^-\pi^0$ events
for one typical experiment in the ensemble~\cite{longki_CPC}.}
\label{fig:kpp_dalitz_onefit}
\end{figure}

The mixing parameters are 
$x'' = x\cos\delta^{}_{K\pi\pi} + y\sin\delta^{}_{K\pi\pi}$ and 
$y'' = -x\sin\delta^{}_{K\pi\pi} + y\cos\delta^{}_{K\pi\pi}$, where
$\delta^{}_{K\pi\pi}$ is the strong phase difference between 
$D^0\ra K^+\pi^-\pi^0$ and $\dbar\ra K^+\pi^-\pi^0$ amplitudes
evaluated at $m^{}_{\pi\pi} = M^{}_\rho$. For this study the
strong phase $\delta^{}_{K\pi\pi}$ is fixed to $10^\circ$, 
and the fitted parameters are $x$ and $y$ directly. 
The results are shown in Fig.~\ref{fig:kpp_dalitz_results}. 
The input values are $x = 0.0258$ and $y = 0.0039$, and the 
RMS of the distributions of residuals are $\delta x = 0.057\%$ 
and $\delta y = 0.049\%$. This precision is approximately 
one order of magnitude better than that achieved by 
BaBar~\cite{babar_dkpp_dalitz}. 
The projections of a typical fit (one experiment in the 
ensemble) are shown in Fig.~\ref{fig:kpp_dalitz_onefit}.
In practice, to extract values for $x$ and $y$ requires 
knowledge of the strong phase $\delta^{}_{K\pi\pi}$; this 
can in principle be measured at BESIII using \cp-tagged 
$D^{}_{CP}\ra K^+\pi^-\pi^0$ decays. 

\subsection{\boldmath Time-integrated (direct) \cp\ Violation}

Belle II will have excellent efficiency for reconstructing multi-body 
final states with very small detector-based asymmetries. Thus the
experiment is ideal for searching for time-integrated (direct) \cp\ 
violation in a variety of final states. A listing of $D^0$, $D^+$, 
and $D^+_s$ decay modes that Belle has studied is given in 
Table~\ref{tab:direct_cpv}. The table lists the \cp\ asymmetry 
$A^{}_{CP} = [\Gamma(D^0) - \Gamma(\dbar)]/[\Gamma(D^0) + \Gamma(\dbar)]$
measured by Belle, and the precision expected for Belle II. The 
latter is estimated by scaling the Belle statistical error
($\sigma^{}_{\rm stat}$) by the ratio of integrated luminosities, 
and by dividing the systematic error into those that scale with 
luminosity such as background shapes measured with control 
samples ($\sigma^{}_{\rm syst}$), and those that do not scale 
with luminosity such as decay time resolution due to 
detector misalignment ($\sigma^{}_{\rm irred}$). The overall
error estimate is calculated as $\sigma^{}_{\rm Belle\ II} =
\sqrt{(\sigma^2_{\rm stat} + \sigma^2_{\rm syst})\cdot 
({\cal L}^{}_{\rm Belle}/50{\rm\ ab}^{-1}) + \sigma^2_{\rm irred}}$.
For most of the decay modes listed, the expected uncertainty 
on $A^{}_{CP}$ is~$\simle 0.10\%$.

\begin{table}[hbt]
\begin{center}
\begin{tabular}{l|ccc}  
Mode &  ${\cal L}$ (fb$^{-1}$) &  $A^{}_{CP}$ (\%) & Belle II 50~ab$^{-1}$ \\
\hline
$D^0\ra K^+K^-$          & 976 & $-0.32\,\pm 0.21\,\pm 0.09$ & $\pm 0.03$ \\
$D^0\ra \pi^+\pi^-$      & 976 & $+0.55\,\pm 0.36\,\pm 0.09$ & $\pm 0.05$ \\
$D^0\ra \pi^0\pi^0$      & 966 & $-0.03\,\pm 0.64\,\pm 0.10$ & $\pm 0.09$ \\
$D^0\ra K_S^0\,\pi^0$    & 966 & $-0.21\,\pm 0.16\,\pm 0.07$ & $\pm 0.03$ \\
$D^0\ra K_S^0\,\eta$     & 791 & $+0.54\,\pm 0.51\,\pm 0.16$ & $\pm 0.07$ \\
$D^0\ra K_S^0\,\eta'$    & 791 & $+0.98\,\pm 0.67\,\pm 0.14$ & $\pm 0.09$ \\
$D^0\ra \pi^+\pi^-\pi^0$ & 532 & $+0.43\,\pm 1.30$           & $\pm 0.13$ \\
$D^0\ra K^+\pi^-\pi^0$   & 281 & $-0.60\,\pm 5.30$           & $\pm 0.40$ \\
$D^0\ra K^+\pi^-\pi^+\pi^-$ & 281 & $-1.80\,\pm 4.40$        & $\pm 0.33$ \\
\hline
$D^+\ra \phi\pi^+$    & 955 & $+0.51\,\pm 0.28\,\pm 0.05$   & $\pm 0.04$ \\
$D^+\ra \eta\pi^+$    & 791 & $+1.74\,\pm 1.13\,\pm 0.19$   & $\pm 0.14$ \\
$D^+\ra \eta'\pi^+$   & 791 & $-0.12\,\pm 1.12\,\pm 0.17$   & $\pm 0.14$ \\
$D^+\ra K^0_S\,\pi^+$ & 977 & $-0.36\,\pm 0.09\,\pm 0.07$   & $\pm 0.03$ \\
$D^+\ra K^0_S\,K^+$   & 977 & $-0.25\,\pm 0.28\,\pm 0.14$   & $\pm 0.05$ \\
\hline
$D^+_s\ra K^0_S\,\pi^+$ & 673 & $+5.45\,\pm 2.50\,\pm 0.33$   & $\pm 0.29$ \\
$D^+_s\ra K^0_S\,K^+$   & 673 & $+0.12\,\pm 0.36\,\pm 0.22$   & $\pm 0.05$ \\
\hline
\end{tabular}
\caption{Time-integrated (direct) \cp\ asymmetries measured
by Belle, and the precision expected for Belle~II in 
50~ab$^{-1}$ of data.}
\label{tab:direct_cpv}
\end{center}
\end{table}

\section{\boldmath Leptonic Decays}

The low backgrounds of an $e^+e^-$ experiment allow one to study
purely leptonic $D^-_{(s)}\ra\ell^-\bar{\nu}$ decays. The branching 
fractions of these decays are proportional to either $f^2_{D}|V^{}_{cd}|^2$
or $f^2_{D^{}_s}|V^{}_{cs}|^2$, where $f^{}_D$ and $f^{}_{D^{}_s}$ 
are decay constants, and $V^{}_{cd}$ and $V^{}_{cs}$ are
Cabibbo-Kobayashi-Maskawa (CKM) matrix elements. Belle has 
measured the branching fraction for $D^-_s\ra\ell^-\bar{\nu}$~\cite{zupanc}, 
and inputting the value of $f^{}_{D^{}_s}$ as calculated 
from lattice QCD~\cite{lattice_flag} results in the 
world's most precise determination of 
$|V^{}_{cs}|$. The $D^-_s\ra\ell^-\bar{\nu}$ 
event sample for Belle II will be significantly larger 
than that for Belle, and this will allow for a more 
precise determination of $|V^{}_{cs}|$. In addition, Belle II
should measure $D^-\ra\mu^-\bar{\nu}$ decays, and from this
branching fraction determine $|V^{}_{cd}|$ to $<2\%$ uncertainty.

\begin{figure}[htb]
\hbox{
\includegraphics[width=0.375\textwidth,angle=-90]{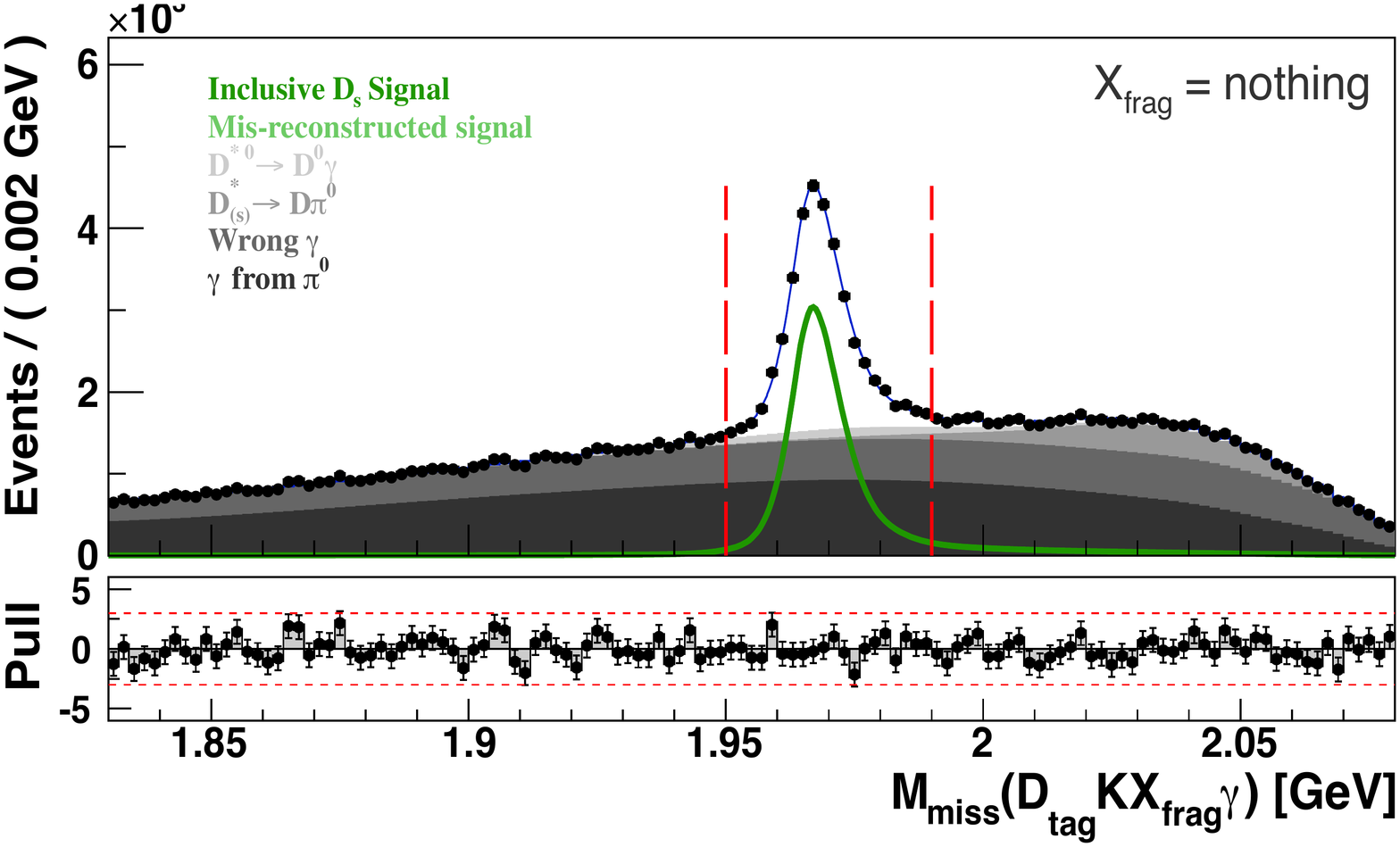}
\hskip0.10in
\includegraphics[width=0.385\textwidth,angle=-90]{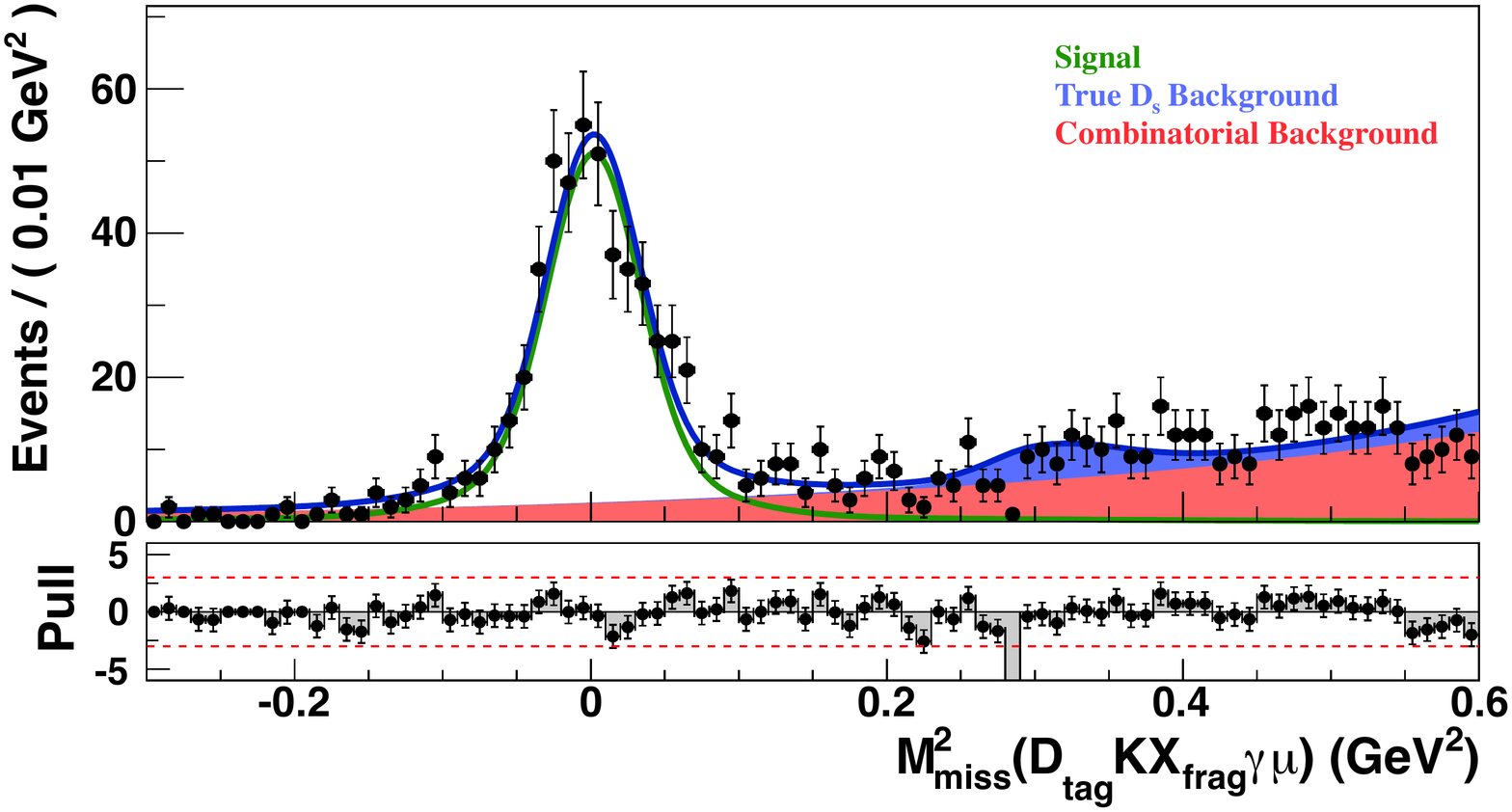}
}
\vskip-0.30in
\caption{The Belle missing mass distributions used 
to measure $D^-_s\ra\mu^-\bar{\nu}$~\cite{zupanc}.
Left: $M^{}_{\rm miss}= \sqrt{(P^{}_{CM}-P^{}_{\rm tag} - 
P^{}_{X^{}_{\rm frag}} - P^{}_{\gamma})^2}$,
which should peak at $M^{}_{D^{}_s}$.
Right: $M^{}_\nu = \sqrt{(P^{}_{CM}-P^{}_{\rm tag} - 
P^{}_{X^{}_{\rm frag}} - P^{}_{\gamma}-P^{}_{\mu^-})^2}$,
which should peak at $M^{}_\nu=0$. }
\label{fig:belle_ds}
\end{figure}

The method used by Belle to reconstruct $D^-_s\ra\mu^-\bar{\nu}$ decays 
is as follows~\cite{zupanc}. First, a ``tag-side'' $D^0$, $D^+$, or 
$\Lambda_c^+$ is reconstructed, nominally recoiling against the signal 
$D^-_s$. The decay modes used for this are listed in Table~\ref{tab:tag_xfrag}.
In addition, tag-side $D^0$ and $D^+$ mesons can be paired with 
a $\pi^+$, $\pi^0$, or $\gamma$ candidate to make a tag-side
$D^{*+}\ra D^0\pi^+$, $D^{*+}\ra D^+\pi^0$, $D^{*0}\ra D^0\pi^0$, 
or $D^{*0}\ra D^0\gamma$ candidate. The remaining pions, kaons, 
and protons in the event are subsequently grouped together into what 
is referred to as the ``fragmentation system'' $X^{}_{\rm frag}$. 
The particle combinations allowed for $X^{}_{\rm frag}$ are also
listed in Table~\ref{tab:tag_xfrag}. Because the signal decay
is $D^-_s$, to conserve strangeness $X^{}_{\rm frag}$ must include 
a $K^+$ or $K^0_S$. If the tag side were a $\Lambda^+_c$, then 
$X^{}_{\rm frag}$ must include a $\bar{p}$ to conserve baryon number.
After $X^{}_{\rm frag}$ is identified, the event is required to 
have a $\mu^-$ candidate nominally originating from 
$D^-_s\ra\mu^-\bar{\nu}$, and a $\gamma$ candidate nominally 
originating from $D^{*-}_s\ra D^-_s\gamma$. The missing mass squared
$M^2_{\rm miss}= (P^{}_{CM}-P^{}_{\rm tag} - P^{}_{X^{}_{\rm frag}} - P^{}_\gamma)^2$
is calculated and required to be within a narrow window centered 
around $M^2_{D_s}$. The signal yield is obtained by fitting 
the ``neutrino'' missing mass distribution 
$M^2_\nu = (P^{}_{CM}-P^{}_{\rm tag} - 
P^{}_{X^{}_{\rm frag}} - P^{}_\gamma -P^{}_{\mu^-})^2$,
which should peak at zero. The missing mass distributions 
for Belle's $D^-_s\ra\mu^-\bar{\nu}$ measurement are shown 
in Fig.~\ref{fig:belle_ds}. The Belle signal yield, and the 
much larger yield expected for Belle II, are listed in 
Table~\ref{tab:dmunu_yields}.

The above method can also be used to search for $D^-\ra\mu^-\bar{\nu}$ 
and $D^0\ra\nu\bar{\nu}$ decays~\cite{ytlai}. In the latter case, 
the $D^0$ is required to originate 
from $D^{*+}\ra D^0\pi^+$, and the daughter $\pi^+$ momentum is used 
when calculating the missing mass. Requiring that $M^{}_{\rm miss}$ 
lie within a narrow window centered around $M^{}_{D^0}$ results in 
an inclusive sample of $D^0$ decays. The Belle yield for this 
sample, and the expected Belle II yield, are also listed 
in Table~\ref{tab:dmunu_yields}. The Belle II yield would 
allow for a $7\times$ more sensitive search for $D^0\ra\nu\bar{\nu}$ 
(or any invisible final state) than that achieved by Belle.

\begin{table}[hbt]
\renewcommand{\arraystretch}{0.95}
\begin{center}
\begin{tabular}{l|c|c|c}  
Tag side: & $D^0$ & $D^+$ & $\Lambda^+_c$ \\
\hline
Decay mode: &
\begin{tabular}{c}
$K^-\pi^+$ \\
$K^-\pi^+\pi^0$ \\ 
$K^-\pi^+\pi^+\pi^-$ \\ 
$K^-\pi^+\pi^+\pi^-\pi^0$ \\ 
$K^0_S\,\pi^+\pi^-$ \\ 
$K^0_S\,\pi^+\pi^-\pi^0$ \\
\end{tabular}
& 
\begin{tabular}{c}
$K^-\pi^+\pi^+$ \\
$K^-\pi^+\pi^+\pi^0$ \\ 
$K^0_S\,\pi^+$ \\ 
$K^0_S\,\pi^+\pi^0$ \\ 
$K^0_S\,\pi^+\pi^+\pi^-$ \\ 
$K^+ K^-\pi^+$ \\ 
\end{tabular} 
& 
\begin{tabular}{c}
$pK^-\pi^+$ \\
$pK^-\pi^+\pi^0$ \\
$p K^0_S$ \\
$\Lambda\pi^+$ \\
$\Lambda\pi^+\pi^0$ \\
$\Lambda\pi^+\pi^+\pi^-$ \\
\end{tabular} 
\\ 
\hline
$X^{}_{\rm frag}:$ & 
\begin{tabular}{c}
$K^0_S\pi^+$ \\
$K^0_S\pi^+\pi^0$ \\
$K^0_S\pi^+\pi^+\pi^-$ \\
$K^+$ \\
$K^+\,\pi^0$ \\
$K^+\,\pi^+\pi^-$ \\
$K^+\,\pi^+\pi^-\pi^0$ \\
\end{tabular} 
& 
\begin{tabular}{c}
$K^0_S$ \\
$K^0_S\,\pi^0$ \\
$K^0_S\,\pi^+\pi^-$ \\
$K^0_S\,\pi^+\pi^-\pi^0$ \\
$K^+\,\pi^-$ \\
$K^+\,\pi^-\pi^0$ \\
$K^+\,\pi^-\pi^+\pi^-$ \\
\end{tabular} 
& 
\begin{tabular}{c}
same as for \\
$D^+$ tag \\
+ $\bar{p}$ \\
\end{tabular} 
\\
\hline
\end{tabular}
\caption{Belle $D^-_s\ra\mu^-\bar{\nu}$ measurement method (see text)~\cite{zupanc}.}
\label{tab:tag_xfrag}
\end{center}
\end{table}

\begin{table}[hbt]
\begin{center}
\begin{tabular}{l|cc}  
 Mode  & Belle yield & Belle II yield (50~ab$^{-1}$)  \\
\hline
$D^-_s\ra\mu^-\bar{\nu}$ & $492\pm 26$ (913~fb$^{-1}$) & 27000 \\
$D^-\ra\mu^-\bar{\nu}$ & $-$ & 1250 \\
$D^0\ra\nu\bar{\nu}$ & $(695\pm 2)\times 10^3$ (924~fb$^{-1}$) & $38\times 10^6$ \\
\hline
\end{tabular}
\caption{Belle's $D^-_s\ra\mu^-\bar{\nu}$~\cite{zupanc} 
and inclusive $D^0$~\cite{ytlai} signal yields, and the 
yields expected for Belle II. The latter are obtained by 
either scaling the Belle results or from MC simulation studies.}
\label{tab:dmunu_yields}
\end{center}
\end{table}

\section{Acknowledgments}

We are grateful to the organizers of CHARM 2016 for a well-organized
workshop in a beautiful location, and for excellent hospitality.


\begin{thebibliography}{99}
\bibitem{charge-conjugates}
Throughout this paper, charge-conjugate modes are implicitly
included unless stated otherwise.
\end{thebibliography}
\end{document}